\documentclass[runningheads,fleqn]{cl2emult}

\usepackage{makeidx}  
\usepackage{graphicx} 
\usepackage{subeqnar} 
\usepackage{multicol} 
\usepackage{cropmark} 
\usepackage{phys}     
\makeindex            

\begin{document}
\title*{Controlling Magneto-Optical Rotation\protect\newline 
via Atomic Coherences}
\toctitle{Controlling Magneto-Optical Rotation via
\protect\newline Quantum Coherences}
%
%
\titlerunning{Controlling Magneto-Optical Rotation}
%
\author{Anil K. Patnaik
\and G. S. Agarwal}
\authorrunning{A. K. Patnaik and G. S. Agarwal}
%
%
\institute{Physical Research Laboratory, Navrangpura,\\ 
Ahmedabad 380 009, India\\
e-mail: aanil@prl.ernet.in}

\maketitle              

\begin{abstract}
An isotropic medium, having magnetic sublevels, when subjected to a magnetic
field or an electromagnetic field can induce anisotropy in the medium; and
as a result the plane of polarization of the probe field can rotate.
Therefore the rotation due to the magnetic field alone, can be {\em controlled 
efficiently} with use of a coherent field. We show, using a control field, 
significant enhancement of the magneto-optical rotation and demonstrate 
the possibility of realizing {\em magneto-optical switch}.
\end{abstract}

\section{Introduction}

An isotropic medium having $m$-degenerate sublevels when subjected to a
magnetic field exhibits birefringence in its response to a polarized optical
field. This is due to the fact that Zeeman splitting of magnetic sublevels
causes asymmetry in the refractive indices for left and right circular
polarization components of the
optical field. The result is magneto-optical rotation (MOR); i.e., the
plane of polarization of the light emerging out of the medium is rotated
with respect to that of the incident. For example consider a $V$-scheme
(say $^{40}Ca$ system) with $4~^1S_0$ as ground state and $4~^1P_1$ as its
excited states, subjected to a magnetic field $\vec{B}$. We probe it by
a linearly polarized light propagating along $\vec{B}$. Let $\chi_+$ and
$\chi_-$ be the susceptibilities corresponding to the right and left circular
polarizations. For small absorption the polarization rotation is given by
\begin{equation}
\theta = \pi k_p l Re(\chi_- - \chi_+),
\end{equation}
where $\vec{k}_p$ corresponds to propagation vector of the probe and
$l$ is the length of the medium.
We note that $\chi_\pm$ depend on the number density of atoms and the
oscillator strength of the transition.

Production of large magneto optical rotation is important for a
number of applications.
In a recent experiment, Sautenkov {\it et al} [1] have shown
enhancement of resonant MOR in an optically dense vapor of {\em Rb}
atom by several order of magnitude. Further, a coherent field can
manipulate the susceptibilities $\chi_\pm$ of the
medium, and in particular can modify the dispersion 
properties of the medium \cite{manipulate}. In another recent experiment 
on coherence 
induced anisotropy, Wielandy and Gaeta \cite{gaeta} have demonstrated 
that when a {\em Rb} vapor cell is illuminated by a
strong laser beam of a particular polarization, and is probed 
by a linearly polarized laser beam, the plane of polarization
of the probe is rotated as the control field induces birefringence in
the medium. 

	In this article, we consider the possibility of control of the 
MOR by using a strong laser beam. We also show that for a chosen 
configuration, inclusion of Doppler effect in the problem gives significant 
enhancement in the MOR - that demonstrates the possibility of realizing 
{\em Magneto-optical switch}.

\section{The Model Scheme and Determination of $\chi_\pm$}
\begin{figure}
\includegraphics[width=.8\textwidth]{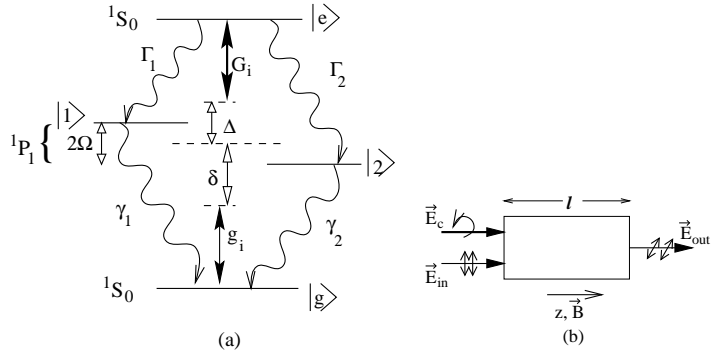}
\caption[]{
(a) The four-level model scheme (say of $^{40}Ca$) having $m$-degenerate
sub-levels as its intermediate states. The symbols in left hand side
denote the energy levels of $^{40}Ca$ atom. $2\Gamma_i$ and $2\gamma_i$ are
the spontaneous decays, $2g_i$ ($2G_i$) is Rabi frequency of the probe (control)
field due to coupling of the intermediate state $|i\rangle$ with $|g\rangle$
($|e\rangle$). The detuning of probe (control) field from the center of
$|1\rangle$ and $|2\rangle$ are represented by $\delta$ ($\Delta$).
$2\Omega$ is the Zeeman split between the intermediate states.
(b) A block diagram that shows the configuration under consideration. $\vec{B}$
defines the quantization axis $z$. The input probe $\vec{E}_{in}$ is
$x$-polarized and the control field is left circularly polarized. Both the
fields propagate along $z$. After passing through the cell, output
is observed through a $y$-polarized analyzer.
}
\label{model}
\end{figure}

We consider a model system [see Fig.\ref{model}] involving say cascade of transitions
$|j=0,m=0\rangle$ (level $|g\rangle$) $\leftrightarrow ~|j=1,m=\pm 1\rangle$
(level $|1\rangle$ and $|2\rangle$) $\leftrightarrow ~|j=0,m=0\rangle$ (level
$|e\rangle$). This for example will be relevant for expressing $^{40}Ca$. The
probe $\vec{E}_p$ will act between the levels $|g\rangle$ and $|1\rangle,
|2\rangle$. We assume in addition the interaction of a control laser
$\vec{E}_c$ to be
nearly resonant with the transition $|e\rangle \leftrightarrow |1\rangle,
|2\rangle$. For simplicity we drop the transition $m=0\leftrightarrow m=0$. We
thus assume the loss to $m=0$ state by spontaneous emission could be pumped
back by an incoherent pump. We derive the density matrix equation for the
system
\begin{equation}
\frac{\partial\rho}{\partial t} = -\frac{i}{\hbar} [{\cal H},\rho] + 
{\rm spontaneous ~decay ~terms},
\label{den-eq}
\end{equation}
that describes the dynamics of the system. Where ${\cal H}$ defines the
Hamiltonian of the system. We solve Eq.(\ref{den-eq}) and could obtain complete 
analytical solutions in the steady state. The susceptibilities of the
medium, to different circularly polarization components, are proportional 
to the off-diagonal density matrix element corresponding to the transition the
polarized field couples. However, here we do not present the complete analytical
results \cite{optcomm}.

	For an $x$-polarized input probe beam, after passing through the
medium, the transmission at the output through a crossed polarized analyzer
(scaled with the input intensity) can be written as
\begin{equation}
T_y = \frac{1}{4}\left| \exp\left( i\frac{\alpha l}{2}\tilde{\chi}_+\right)
-\exp\left( i\frac{\alpha l}{2}\tilde{\chi}_-\right)\right|^2;
\end{equation}
where $\tilde{\chi}_\pm$ represent the normalized susceptibilities.
In rest of this article, we drop the tilde for brevity. The quantity
$\alpha l$ gives resonant absorption of the medium.
For a particularly interesting case of a $\sigma_-$ polarized control field 
(i.e., ${\cal E}_{c+} 
= 0$, ${\cal E}_{c-}\ne 0$), $\chi_\pm$  are found to be
\begin{equation}
\chi_- \equiv \frac{i\gamma}{{\bf (}\gamma +i(\delta -\Omega){\bf )}},
\label{chim}
\end{equation}
\begin{equation}
\chi_+ \equiv \frac{i\gamma{\bf (}\Gamma_1 +\Gamma_2 + i(\Delta + \delta){\bf )}
}
{|G_1|^2 + {\bf (}\gamma +i(\delta+\Omega){\bf )(}\Gamma_1 +\Gamma_2 + i(\Delta
+ \delta){\bf )}},
\label{chip}
\end{equation}
where the symbols represent the parameters defined in Fig.1. For simplicity,
we have assumed $\gamma_i = \gamma$. In all the plots, all
frequencies are scaled with $\gamma$ ($=\Gamma_i$).
Clearly, $\chi_-$ is independent of control field parameters, where as
$\chi_+$ depends strongly on the strength and frequency of the control
field. For large $|G_1|$, both real and imaginary part of $\chi_+$
will show Autler-Townes splitting and therefore will cause a large 
asymmetry between the two polarization components.

\section {Laser Field Induced Enhancement of MOR}
\begin{figure}
\includegraphics[width=.9\textwidth]{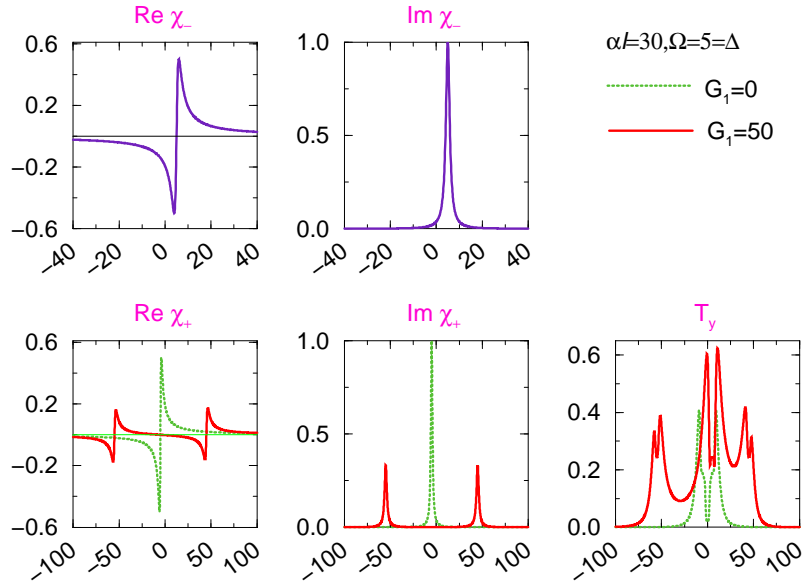}
\caption[]{
Large enhancement of MOR for a range of frequencies of the probe beam
is shown. The parameters are mentioned above. 
}
\label{fig2}
\end{figure}

From Eq.(\ref{chim}) and (\ref{chip}), we note that, 
$\chi_+ (\Omega = 0) \ne \chi_+ (\Omega =0)$ for $G_c \ne 0$. 
Therefore the birefringence can be induced 
in the medium by the laser field even in absence of the magnetic field 
\cite{gaeta}.
Thus one observes a large rotation of the polarization of the probe. 
There are many reports of laser field induced birefringence \cite{others}
which had suffered from large absorption, resulting in a very small rotation
signal at the output. Further when magnetic field is 
present, {\em new frequency regions} are created by application of the control 
field 
where significant enhancement of MOR signal is obtained 
particularly in the 
regions where the MOR, otherwise, is negligible. For example in 
Fig.\ref{fig2}, at $\delta \sim \pm 50$, there is a large enhancement of MOR. 
For detuned control fields the 
MOR could be enhanced further and we also analyze the case 
of an elliptically polarized control field and identify many interesting 
parameter domains where large MOR is obtained (results not shown here).

\noindent
\section {Realization of a Magneto-Optical Switch}

We next consider a Doppler broadened medium where one needs to average
$\chi_\pm$ over the atomic velocity distribution function inside the cell. 
We have identified a configuration that {\em takes 
advantage of the Doppler broadening} to increase the asymmetry between 
$\chi_\pm$, and hence to obtain 
significantly large enhancement of the MOR signal.
\begin{figure}
\includegraphics[width=.75\textwidth]{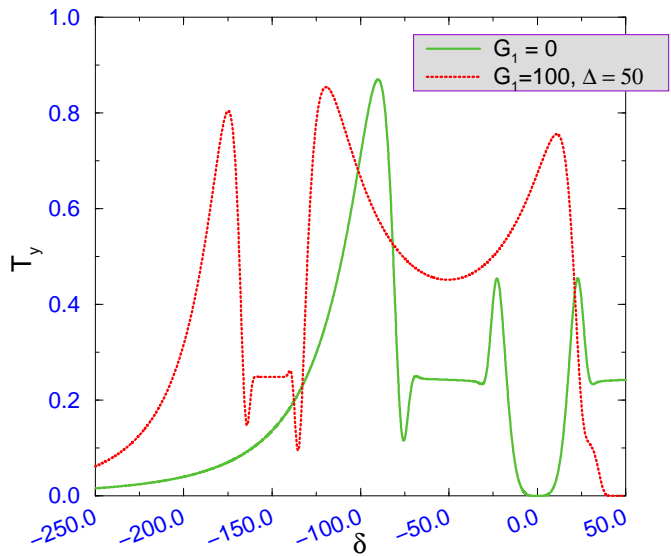}
\caption[]{
Significant enhancement of MOR in a Doppler broadened medium. The
large enhancement factor at $\delta = 0$ demonstrates the possibility of
{\em Magneto-optical switch}. Here the parameters used are 
$\alpha  l = 300$ and $\Omega = 50$.  
}
\label{fig3}
\end{figure}

We consider the same configuration as in Fig.1 but with 
control field ($\sigma_-$ polarized) and the probe field ($x$-polarized)
counter propagating to each other. The 
$\sigma_-$ component of the probe is, thus, Doppler broadened as it does
not see the control field, but on the other hand $\sigma_+$ component 
experiences the counter propagating $\sigma_-$ polarized control field and, 
therefore, is almost 
Doppler free. That leads to enhancement of the asymmetry between $\chi_+$ and 
$\chi_-$. 
We have derived analytical expressions for the Doppler averaged values
of $\chi_\pm$ and hence for the rotation spectra. In Fig.\ref{fig3},
at $\delta \sim 0$, MOR enhancement factor is as large as 
$3.7\times 10^4$, compared to that of MOR with no control field case. 
Such an action of control field can be used as a {\em magneto-optical 
switch that switches the given polarization state of the probe field 
to its orthogonal component}. Several 
interesting set of parameters are identified where the $y$-polarized signal
intensity at the output is as large as $\sim 92\%$ of the $x$-polarized 
input intensity.

\section{conclusion}
	We have shown how a control field can induce birefringence and 
enhance MOR. We have also shown that it can create new regions where MOR
enhances significantly. In an inhomogeneously broadened medium, we 
have shown that, the control field enhances MOR to a large order of magnitude, 
and therefore it demonstrates the possibility of realizing 
a {\em magneto-optical switch}.


\end{document}